\documentclass[12pt,a4paper]{article}
\usepackage{geometry,amsmath,amsfonts,amssymb,graphicx}
\begin{document}

\title{The Stability of Branonium}

\author{James Ellison\footnote{email: j.b.ellison@sussex.ac.uk}\; and
        Andr\'e Lukas\footnote{email: a.lukas@sussex.ac.uk}\\\\
        Department of Physics and Astronomy, University of Sussex,\\
        Falmer, Brighton BN1 9QJ, United Kingdom}

\maketitle
\thispagestyle{empty}

\begin{abstract}

\noindent
We analyse the orbital motion of a light anti D6-brane in the
presence of a stack of heavy, distant D6-branes in ten dimensions,
taking account of possible time-variations in the background moduli
fields. The Coulomb-like central potential arising through
brane-antibrane interactions is then modified to include
time-dependent prefactors, which generally preclude the existence of
stable elliptical orbits.

\end{abstract}

\newpage

\section*{Introduction}

In recent years, it has become clear that the construction of realistic or semi-realistic
particle physics models from string- or M-theory typically involves D- or M-branes as part
of the vacuum configuration. Frequently, these branes have moduli on their worldvolume which
allow brane motion in the transverse internal space or some part thereof. When considering
the cosmology of such models, motion of these branes is therefore an interesting generic
feature which deserves systematic study~\cite{Dvali:1999pa}--\cite{Copeland:2002fv}. The
purpose of this paper is to analyse time-evolution of branes in the framework of a
particular set of models following Ref.~\cite{Burgess:2003qv}, focusing on the interplay of
brane motion and evolution of gravitational moduli, an aspect often neglected in this
context. Our main points will be, firstly, that in a realistic situation the gravitational
moduli cannot be truncated off consistently and, therefore, necessarily evolve in time
whenever the brane moves and, secondly, that this significantly affects the behaviour of the
system.

More specifically, we will consider the dynamics of a light IIA or IIB anti D$p$-brane in
the background of a heavy D$p$-brane. For the case of an anti D6-brane, it was shown in
Ref.~\cite{Burgess:2003qv}, that the worldvolume action in the background of a stack of D6
branes has rather unusual properties. In particular, the fully relativistic, non-linear
action allows for the existence of non-precessing, elliptical orbits for the D6 anti-brane
similar to those admitted for a non-relativistic particle experiencing a central potential.
The stability of these orbits against tidal forces and radiation emission was then computed,
with the conclusion that both effects could be minimised at large distances. This
``planetary'' system of branes was named ``branonium'' in Ref.~\cite{Burgess:2003qv}. Its
existence seems to indicate that extended objects can orbit around one another for a
considerable period of time, perhaps with interesting cosmological effects if, for example,
the central stack in this setup supports our standard-model universe. However, it should be
noted that the analysis of Ref.~\cite{Burgess:2003qv} was carried out in a static background
geometry.

It is known from related systems~\cite{Copeland:2001zp} that the inclusion of gravitational
moduli, such as internal radii, is necessary for consistency and qualitatively alters the
evolution of the fields . We are, therefore, asking how the inclusion of gravitational
moduli affects the above-mentioned system of D$p$ branes and, in particular, the orbital
motion found in the D6 case. It is conceivable that even a small interaction between brane
position moduli and gravitational moduli prevents the motion from being cyclic and leads
to a qualitatively different behaviour, with the deviation from the pure ``on-the-brane''
result accumulating with each revolution.

To answer this question, we will first derive the effective $d=p+1$ dimensional theory
including the brane- as well as the bulk-moduli, by reducing on the (compactified)
transverse space. We then specialise to the $p=6$ case and study the evolution of the
anti-D6 brane including the dilaton and the scale factor of the transverse space.


\section*{Bulk action and $p$-brane solutions}

To set up the framework and introduce our notation we start by reviewing the bulk action and
the standard form of its $p$-brane solutions, largely following
Ref.~\cite{Duff:1994an,Stelle:1998xg}. Consider the following action in $D$ dimensions
\begin{align} \label{bulkaction}
S_{Bulk}= \frac{1}{2\kappa^{2}}\int d^{D}x \sqrt{-g}
\left[R(g)-\frac{1}{2}\nabla_{A}\Phi\nabla^{A}\Phi
-\frac{1}{2(d+1)!}e^{a\Phi}F^{2}_{[d+1]}\right]
\end{align}

where $g$ is the Einstein-frame metric, $\Phi$ is the dilaton, $F_{[d+1]}$ is the $d+1$-form
field strength for a \mbox{$d$-form} potential $A_{[d]}$, and $a$ is a constant. Of course,
we have in mind that $D=10$ and that the above action represents a sub-sector of IIA or IIB
supergravity; we will state below the precise conditions for this to be the case. One can
look for supersymmetric BPS solutions to the above action where the fields only depend on
the isotropic radial coordinate transverse to a flat, $d$-dimensional hyperplane embedded in
the full space-time. This plane is then taken to represent the worldvolume ``history'' of an
extended object with $p=d-1$ spatial dimensions - namely a $p$-brane - whose presence
dictates a precise variation of the bosonic fields $g,\Phi,F$ transverse to its spatial
extent. Much like the extreme Reissner-Nordstrom black-hole solution of $4D$ supergravity,
these $p$-branes have their tensions $T$ equal to their charges $Q$ and preserve one-half of
the supersymmetries found for a Minkowski background.

Splitting the space-time coordinates into the two sets as $(x^{A})=(x^{\mu},y^{m})$, where
$\mu=0,..,p$ and $m=d,..,D-1$, an elementary, ``electric'' $p$-brane solution to the
action~\eqref{bulkaction} is given by \cite{Duff:1994an}
\begin{align} \label{ansatz}
ds^{2} &= e^{2\nu_{0}}h^{-\tilde{\gamma}}\eta_{\mu\nu}dx^{\mu}dx^{\nu} +
e^{2\beta_{0}}h^{\gamma}\delta_{mn}dy^{m}dy^{n}\\
e^{\Phi} &= e^{\phi_{0}}h^{\sigma} \\
A_{\mu_{1} \ldots \mu_{d}} &= \epsilon_{\mu_{1} \ldots \mu_{d}} \: e^{d\nu_{0}
-\frac{1}{2}a\phi_{0}}\zeta h^{-1}\; ,
\end{align}

where arbitrary integration constants $\nu_{0},\beta_{0},\phi_{0}$ have been retained for
later promotion to moduli fields, and
\begin{align*}
\gamma = \frac{4d}{\Delta(D-2)}, \quad \tilde{\gamma}=\frac{4\tilde{d}}{\Delta(D-2)}, \quad
\sigma = \frac{2a}{\Delta}, \quad \zeta = \frac{2}{\sqrt{\Delta}}, \quad \Delta = a^2+
\frac{2d\tilde{d}}{D-2}
\end{align*}

with $\tilde{d}=D-d-2$. The harmonic function $h$ is given by
\begin{align*}
h = 1+\frac{k_{0}}{r^{\tilde{d}}}\;e^{-\frac{1}{2}a\phi_{0}-\tilde{d}\beta_{0}}
\end{align*}
where $r$ is the isotropic radial coordinate defined by
\begin{align*}
r^{2}=\delta_{mn}y^{m}y^{n}\; .
\end{align*}
Further, $k_{0}$ is a dimensionful constant whose value is fixed in terms of the positive tension
$T$ of the $p$-brane source.

Given their importance in the branonium analysis, it will also prove useful to identify a
`stringy' subset of these solutions for which $D=10$, $\Delta=4$,
$2\sigma=a=(4-d)/2$. For these special choices the action \eqref{bulkaction} is a truncated,
low-energy description of a type II string theory, with $F_{[d+1]}$ an RR form field, so
there is an alternative D-brane description of these $p$-branes. Specifically, any such
$p$-brane can be reinterpreted as a collection of coincident D$p$-branes - with $k_{0}$
fixed by the total mass density of the D$p$'s - provided that (in the Einstein frame)
\begin{align}\label{kcond}
k \equiv k_{0}e^{-\frac{1}{2}a\phi_{0}} \gg 1\; .
\end{align}
This condition ensures that the underlying D$p$-brane collection always give rise to an
object with a characteristic length-scale much larger than the string length, which is a
necessary feature of a supergravity $p$-brane.

Before proceeding further, we emphasise that the entire set of Einstein-frame solutions
listed in \eqref{ansatz} can be reached from the solutions used in \cite{Burgess:2003qv} by
simple coordinate transformations, meaning that no new features have been added to these
solutions. Once allowance has been made for the Einstein/string frame distinction, the
condition~\eqref{kcond} is also exactly as stipulated in \cite{Burgess:2003qv}.

\section*{Branonium}
In Ref.~\cite{Burgess:2003qv} it was shown how a single ``probe'' anti D6 brane could, in
ten dimensions, execute stable orbital motion in the background \eqref{ansatz} created by a
large collection of coincident D6-branes. This configuration, by analogy with the
electron-positron system positronium, was dubbed ``branonium''. We now briefly re-derive the
probe action presented in \cite{Burgess:2003qv}. The \mbox{$d$-dimensional} worldvolume
action (in the Einstein frame) for a general probe is given by
\begin{align} \label{braneaction}
S_{probe} = -T\int d^{d}\xi\sqrt{-\hat{g}} \:e^{-\frac{1}{2}a\hat{\Phi}} + Q\int
\hat{A}_{[d]}
\end{align}
where $T$ is the bare positive tension of the probe, $Q$ is its bare charge and $\xi^{\mu}$
are intrinsic coordinates on the worldvolume. The quantities
$\hat{g},\hat{\Phi},\hat{A}_{[d]}$ are simply pullbacks of the corresponding bulk fields:
\begin{align*}
\hat{\Phi}=\Phi(X)\quad, \quad \hat{g}_{\mu\nu}= \frac{\partial X^{A}}{\partial \xi^{\mu}}
\frac{\partial X^{B}}{\partial \xi^{\nu}} g_{AB}(X) \quad, \quad
\hat{A}_{\mu_{1}\ldots\mu_{d}}=\frac{\partial X^{\nu_{1}}}{\partial \xi^{\mu_{1}}}\ldots
\frac{\partial X^{\nu_{d}}}{\partial \xi^{\mu_{d}}}\: A_{\nu_{1}\ldots \nu_{d}}(X)
\end{align*}
where the $X^{A}$ are embedding coordinates that locate the probe in the ambient space-time.
Exploiting the $\xi^{\mu}$ coordinate reparameterisation freedom on the probe worldvolume,
one can choose a gauge such that the $X^{A}$ become
\begin{align*}
X^{\mu}=\xi^{\mu} \qquad, \qquad X^{m}=Y^{m}(\xi^{\mu})
\end{align*}
leaving $Y^{m}$ as the physical degrees of freedom that locate the probe in the space
transverse to the source. Inserting the background solutions \eqref{ansatz} into the probe
action, and momentarily setting $\nu_{0}=\beta_{0}=0$, one finds the result
\begin{align}
S_{probe} = -\mathcal{M} \int d^{d}x \left[ h^{-\eta}\sqrt{1+h^{\omega}\sum_n(\partial
Y^{n})^{2}} - q\zeta h^{-1}\right]
\end{align}
with
\begin{align*}
\eta = \frac{2[a(\tilde{d}-d)+\tilde{d}d]}{\Delta(D-2)}, \qquad \omega = \frac{4}{\Delta},
\qquad q=\pm1, \qquad \mathcal{M}=Te^{-\frac{1}{2}a\phi_{0}}
\end{align*}
where $\mathcal{M}$ corresponds to the physical ADM mass density of
the probe measured at infinity where there is a constant dilaton vev
$\phi_{0}$. This concurs with the results of Ref.~\cite{Burgess:2003qv}. In
the special case of a probe anti D6 brane circling a stack of source
D6's in ten dimensions, one finds $\eta=\omega=\zeta=\tilde{d}=-q=1$
and the function $h$ becomes a central potential. The fully
relativistic worldvolume Lagrangian for the probe motion then mimics
that of a non-relativistic bound particle experiencing an inverse
square law force of attraction.

\section*{Time-dependent backgrounds}

We now proceed to include the gravitational moduli in the calculation. In our specific case,
these moduli are the zero mode $\phi_0$ of the dilaton and the scale factor $\beta_0$ of the
transverse space (as well as a $d$-dimensional conformal factor $\nu_0$). They have already
been included in the above $p$-brane solution and we reiterate that Eq.~\eqref{ansatz}
constitutes a solution of the action~\eqref{bulkaction} for all moduli values. We now derive
an effective $d$-dimensional theory that includes these gravitational moduli fields as well
as the position moduli $Y^{m}(\xi)$ of the anti-brane. To this end, we promote $\nu_0$,
$\phi_0$ and $\beta_0$ to fields $\nu$, $\beta$ and $\phi$, depending on the longitudinal
coordinates $x^\mu$, and modify the solution~\eqref{ansatz} to
\begin{align} \label{newansatz}
ds^{2} &= e^{2\nu}h^{-\tilde{\gamma}}\tilde{g}_{\mu\nu}(x)dx^{\mu}dx^{\nu} +
e^{2\beta}h^{\gamma}\delta_{mn}dy^{m}dy^{n}\\
e^{\Phi} &= e^{\phi}h^{\sigma} \\
A_{\mu_{1} \ldots \mu_{d}} &= \epsilon_{\mu_{1} \ldots \mu{d}} \: e^{d\nu
-\frac{1}{2}a\phi}\zeta h^{-1}\; .
\end{align}
Here, $\tilde{g}$ will be the $d$-dimensional Einstein-frame metric (after some
appropriate choice of the conformal factor $\nu$ to be specified later), and the harmonic function
\begin{align*}
h = 1+\frac{k_{0}}{r^{\tilde{d}}}\;e^{-\frac{1}{2}a\phi-\tilde{d}\beta}
\end{align*}
is now $x^\mu$-dependent. Ideally, we would like to insert this solution into the action
\eqref{bulkaction} and explicitly integrate out the $y$ dependence, leaving a
$d$-dimensional covariant bulk action that combines with the $d$-dimensional worldvolume
action~\eqref{braneaction} of the anti brane. The total, effective action could then be
solved for the coupled behaviour of the bulk and brane moduli.

However, due to the singularity of $h$ at $r=0$, and the fact that the transverse dimensions
are non-compact, any integration over $y$ is likely to be divergent. Worse, each term in the
action will generally pick up different powers of the function $h$ such that the divergences
cannot even be gathered into a common divergence multiplying the entire action. This means
that divergences will proliferate in the equations of motion, and our effective theory will
be poorly defined. As a first step toward rectifying this problem, we assume there is some
region of moduli space in which powers of $h$ are well approximated by a series-expansion to
first order:
\begin{align*}
h^{n} = 1 + \frac{n\epsilon}{r^{\tilde{d}}} + O(\epsilon^{2})
\end{align*}
where
\begin{align*}
\epsilon=k_{0}e^{-\frac{1}{2}a\phi-\tilde{d}\beta} \ll 1
\end{align*}
defines that portion of moduli-space for which the expansion is valid. Since we know
from~\eqref{kcond} that $k\gg1$ in the supergravity regime, and we assume $\tilde{d}>0$ in
what follows, this necessitates taking $\beta \gg1$. This is desirable in any event, since
the antibrane must be at a large proper distance from the source at $r=0$ in order to
suppress open-string tachyon instabilities. So we expect this expansion to be sensible, and
to be equivalent to the non-relativistic expansion in powers of $k/r^{\tilde{d}}$ used in
the original branonium analysis.

The point of the expansion is that it allows us to cleanly separate our problems into two
categories: divergences associated with the constant $\epsilon^{0}=1$ term, and divergences
associated with the linear $\epsilon$ term. We deal with the linear term first, by noting
that there is always some arbitrariness when defining the $d$-dimensional moduli fields. In
particular, we are free to redefine the modulus $\phi$ such that it measures fluctuations
above some small, non-zero background value; that is, $\phi$ is no longer exactly equivalent
to the asymptotic value of the $D$-dimensional dilaton $\Phi$. The net effect of this change
is to adjust the harmonic function $h$ to the form:
\begin{align*}
h = 1 + \epsilon \left( \frac{1}{r^{\tilde{d}}}-c_{1} \right)
\end{align*}
where $c_{1}$ is a constant. In addition, we introduce some large cut-off $r_c$ for the integration
over the transverse space, so that its coordinate volume is given by  $V \sim r_{c}^{\tilde{d}+2}$.
Then choosing the constant $c_1$ as
\begin{align*}
c_{1} = \left(\frac{\tilde{d}+2}{\tilde{d}+1}\right)\frac{1}{r_{c}}
\end{align*}
we automatically cancel the integral of the $1/r^{\tilde{d}}$ term. Moreover, in the
non-compact limit $r_{c}\rightarrow \infty$ we find that $c_{1}$ is infinitesimally small,
and so has a negligible impact in the remaining effective action. Overall, therefore, terms
in the bulk action that are linear in $\epsilon$ can always be made to \emph{vanish} after
the integration.

The contribution of the leading term in $h$ is now also regulated by the finite cutoff
$r_c$. After integrating over the transverse space, the reduced bulk action becomes
proportional to the volume $V$. In the infinite volume limit, $V\rightarrow\infty$, this
implies a decoupling of the bulk action and, hence, of the gravitational moduli from the
brane action. In this limit, therefore, it is consistent to set the gravitational moduli to
constants. However, as we will see later, even in this case small initial velocities for the
bulk moduli will affect the orbital motion of the probe brane. More interesting, and more
realistic from the point of view of brane-world model building, is the case of a compact
transverse space with finite volume $V$. For simplicity, we consider toroidal (or
orientifold) compactifications. Of course, a compact transverse space requires the
introduction of negatively charged orientifold $p$-planes so that the total RR charge in the
compact space is zero. The resulting D$p$-O$p$ system can be described by a solution of the
type~\eqref{ansatz} but with the harmonic function $h$ modified to
\begin{align}\label{trueh}
h = 1 + \epsilon \left( \frac{1}{r^{\tilde{d}}}-c_{1} \right) - \epsilon
\left(\frac{1}{|\textbf{r}-\textbf{r}_{0}|^{\tilde{d}}}-c_{2}\right) + \text{images}
\end{align}
where $\textbf{r}_{0}$ is the position of the O$p$ stack, and the inclusion of orientifold
images of the D$p$-branes has been indicated. These images are a necessary feature of the
periodic identifications. We are now assuming a sufficiently large transverse volume and a
trajectory of the probe brane which is far away from the orientifold planes. In this case,
the effect of the orientifold planes and the images on the anti-D$p$ brane can be neglected;
that is, we can still think of the anti-D$p$ brane as essentially moving in the background
generated by the heavy D$p$-brane alone. Notice as well that the influence of both the brane
and orientifold plane is controlled by the same expansion parameter $\epsilon$. This
guarantees that, once we have arranged the effect of the orientifold plane and the images on
the anti-brane to be negligible, this will remain true throughout bulk moduli space.

To summarise our procedure, we will use $h$ in the form \eqref{trueh}, insert the ansatz
\eqref{newansatz} into the bulk action, expand all powers of $h$ up to linear order in
$\epsilon$, choose the $c_{i}$ such that these linear order terms integrate exactly to zero,
and then factor out $\epsilon^{0}$ to give a large but \emph{finite} volume $V$. The bulk
action so obtained is formally the same as if $h=1$ had been assumed from the start, the
difference arising only in the probe action where the pullback of the function $h$ is felt
explicitly. We will then take the probe's average radial position $R$ to be far from the
position of the orientifolds, and so discount the images and the radially-dependent term in
$h(R)$ due to the orientifold. As there is no integration over $y$ in the anti-brane action,
we will then dispense with the constants $c_{i}$, since they are still small enough to have
a negligible impact upon the dynamics.

The resulting effective action is explicitly given by
\begin{align}
\begin{split}\label{bulkeffective}
S =& \frac{V}{2\kappa^{2}} \int d^{d}x \sqrt{-\tilde{g}} \left\{ \tilde{R}
-\frac{1}{2}(\partial\phi)^{2}-\frac{(\tilde{d}+2)(\tilde{d}+d)}{(d-2)}(\partial \beta)^2
\right. \\
 -&\frac{2\kappa^{2} T}{V} \left[ \frac{1}{2}h^{\omega-\eta}e^{-\frac{1}{2}a\phi-\tilde{d}\beta}
\sum_n(\partial
Y^{n})^2 + e^{-\frac{1}{2}a\phi-\frac{d(\tilde{d}+2)}{(d-2)}\beta} \left(h^{-\eta}-q \zeta
h^{-1}\right) \right] \left. \vphantom{\int}\right\}
\end{split}
\end{align}
where the convenient choice
\begin{align*}
(d-2)\nu + (\tilde{d}+2)\beta = 0
\end{align*}
has eliminated $\nu$ and has guaranteed a canonical Einstein-Hilbert term in $d$ dimensions
for the Ricci scalar $\tilde{R}$ built out of $\tilde{g}$. Notice as well that the probe
worldvolume action has been linearised to at most two derivatives, which is consistent with
the assumption that $\epsilon \ll 1$.

Note that the kinetic terms for the brane position moduli $Y^n$ are non-trivial, depending
on exponentials of the bulk moduli $\phi$ and $\beta$. This means that the bulk moduli
equations of motion have source terms proportional to the kinetic energy of $Y^n$ and the
inverse volume $1/V$ of the transverse space. Hence, whenever the brane moves and the volume
$V$ is kept finite the gravitational moduli will also evolve in time even if they are static
at some initial time. Only in the decompactification limit $V\rightarrow \infty$ can the
bulk moduli $\phi$ and $\beta$ be set to constants consistently. However, even in this case
non-vanishing velocities for the bulk moduli will affect the evolution of the brane due to
the non-trivial $Y^n$ kinetic terms. We will study both cases, the case of finite $V$ and
initially static bulk moduli, and the case $V\rightarrow \infty$ with small initial
velocities for $\phi$ and $\beta$, in detail below.

We now focus on the case of an anti D6-brane in the background of a stack of D6 branes in
ten dimensions, with parameters
\begin{align*}
a=-3/2,  \quad \sigma=-3/4, \quad \Delta=4, \quad \zeta=\omega=\eta=\tilde{d}=-q=1\; .
\end{align*}
The above action then specialises to
\begin{align}
S = \frac{V}{2\kappa^{2}} \int d^{7}x \sqrt{-\tilde{g}} \left\{ \tilde{R}
-\frac{1}{2}(\partial\phi)^{2}-\frac{24}{5}(\partial \beta)^2 -\frac{2\kappa^{2}T}{V} \left[
\frac{1}{2}e^{\frac{3}{4}\phi-\beta}\sum_{n=1}^{3}(\partial Y^{n})^2 +
2e^{\frac{3}{4}\phi-\frac{21}{5}\beta}h^{-1} \right] \right\}\; .
\end{align}
Here $h$ is given by
\begin{align*}
h = 1 + \frac{\epsilon}{R}
\end{align*}
where $R^{2} = \delta_{mn}Y^{m}Y^{n}$ is the radial coordinate locating the antibrane.
The above action is computed to first order in the parameter
\begin{align*}
\epsilon=k_{0}e^{\frac{3}{4}\phi-\beta}
\end{align*}
and provides a valid description as long as $\epsilon$ is small, the distance $R$ from
the central D6-brane is sufficiently large and $R$ is much smaller than $V^{1/3}$, the
typical size of the transverse space.


\section*{Cosmology}

We now assume that all moduli are
time-dependent only, and adopt a simple FRW ansatz for $\tilde{g}$ of the form
\begin{align*}
d\tilde{s_{7}}^{2} = -dt^{2}+e^{2\alpha}\delta_{ij}dx^{i}dx^{j}\; .
\end{align*}
Here $\alpha\equiv\alpha(t)$ denotes the scale-factor for the spatial
part of the metric $\tilde{g}$  and $i,j=1,..,6$ index the corresponding coordinates.
Then the Einstein equations are given by
\begin{align}
-15\dot{\alpha}^{2} + \frac{1}{4}\dot{\phi}^2 + \frac{12}{5}\dot{\beta}^{2} +
\frac{2\kappa^{2}T}{V}\left[\frac{1}{4}\sum_{n=1}^{3}e^{\frac{3}{4}\phi-\beta}(\:\dot{Y}^{n})^{2}
+e^{\frac{3}{4}\phi-\frac{21}{5}\beta}\left(1 - \frac{k_{0}e^{\frac{3}{4}\phi-\beta}}{R} \right)\right] &=0\label{Friedmann}\\
5\ddot{\alpha}+15\dot{\alpha}^{2}+ \frac{1}{4}\dot{\phi}^2 + \frac{12}{5}\dot{\beta}^{2} +
\frac{2\kappa^{2}T}{V}\left[\frac{1}{4}\sum_{n=1}^{3}e^{\frac{3}{4}\phi-\beta}(\:\dot{Y}^{n})^{2}
- e^{\frac{3}{4}\phi-\frac{21}{5}\beta}\left(1 - \frac{k_{0}e^{\frac{3}{4}\phi-\beta}}{R}
\right) \right] &=0
\end{align}
while the moduli equations of motion read
\begin{align}
\ddot{\phi} +6\dot{\alpha}\dot{\phi}
-\frac{2\kappa^{2}T}{V}\left[\frac{3}{8}\sum_{n=1}^{3}e^{\frac{3}{4}\phi-\beta}(\:\dot{Y}^{n})^{2}
-\frac{3}{2}e^{\frac{3}{4}\phi-\frac{21}{5}\beta}\left(1+\frac{2k_{0}e^{\frac{3}{4}\phi-\beta}}{R}\right)\right]&= 0\\
\ddot{\beta} + 6\dot{\alpha}\dot{\beta}
+\frac{2\kappa^{2}T}{V}\left[\frac{5}{96}\sum_{n=1}^{3}e^{\frac{3}{4}\phi-\beta}(\:\dot{Y}^{n})^{2}
-\frac{7}{8}e^{\frac{3}{4}\phi-\frac{21}{5}\beta}\left(1+\frac{26k_{0}e^{\frac{3}{4}\phi-\beta}}{21R}\right)\right]  &= 0 \\
\ddot{Y}^{n}+\left(6\dot{\alpha}+\frac{3}{4}\dot{\phi}-\dot{\beta}\right)\dot{Y}^{n}
+2k_{0}e^{\frac{3}{4}\phi-\frac{21}{5}\beta}\frac{Y^{n}}{R^{3}} &=0\; .
\end{align}
Here we have linearised $h$ to first-order in $\epsilon$, consistent with our earlier
approximation. These equations can now be numerically solved to determine the behaviour of
the probe, provided that we explicitly fix the free parameters $V$ and $k_{0}$. The
constants $\kappa$ and $T$ are predetermined in terms of the string length $l_{s}$ as
\begin{align*}
T = \frac{\sqrt{\pi}}{\kappa}(2\pi l_{s})^{-3}, \qquad 2\kappa^{2} = (2\pi)^{7}l_{s}^{8}\; .
\end{align*}
In the following we will set $l_{s}$ to one, for simplicity, which implies that
\begin{align*}
\frac{2\kappa^{2}T}{V} = \frac{2\pi}{V}
\end{align*}
Then $V$ measures the transverse coordinate volume in units of $l_{s}^3$, while $R$ measures
the distance of the anti D6-brane from the central stack in units of $l_{s}$. With this
choice of length unit we must also take $k_{0}\gg1$ in order for the supergravity analysis
to be valid. Explicit choices for these parameters now crucially depend on maintaining the
antibrane at a distance far enough from the branes to suppress tachyonic effects, whilst not
allowing it to sense either the orientifold or the periodic geometry of the transverse
space. That is, we wish the typical antibrane distance from the the central stack to be
larger than one in string units and, at the same time, we want this distance to be much
smaller than the size of the compact space. Provided that we always choose $\beta \gg 1$
(i.e. $\epsilon \ll 1$) it is sufficient that these conditions are met for the coordinate
values of distances and sizes. In other words, we will require $R$ to be much larger than
one and much smaller than $V^{1/3}$. Typical values will be $R_0=100$ for the initial radius
$R_0$ and $V \sim (10R_{0})^{3}=10^{9}$ for the coordinate volume of the transverse space.
The antibrane can then orbit without undue influence from either the orientifolds or the
images, provided that it does not approach $R \sim 1000$.

Bearing this in mind, we make the following choices. We assume that $k_{0}=100$ and $k=10$
such that $\phi_{0}=4/3\ln{(0.1})=-3.1$\:. This is compatible with the fact that the string
coupling $g_{s}=e^{\phi_{0}}$ should satisfy $g_{s}\ll 1$ in order to avoid large
string loop corrections. Remembering that we require a small initial value $\epsilon_{0} \ll
1$ for the expansion parameter $\epsilon$, we then set $\epsilon_{0}=0.01$ such that $
\beta_{0}=\ln{10^4}=9.2$. For the antibrane moduli we assume that initially
$Y_{3}=\dot{Y}_{3}=0$ so that there is some chance of orbital motion in the $(Y^{1},Y^{2})$
phase plane. We then choose $Y_{1,2}$ and their time derivatives such that $Y_{1}=100,
Y_{2}=0$ and $\dot{Y}_{1}=0, \dot{Y}_{2}=10^{-9}$. This initially sets $R_{0}=100$ such that
we should choose $V=(1000)^{3}=10^{9}$; the value of $\dot{Y}^{2}$ is merely to ensure that
as $V$ becomes infinitely large the system approaches one of the exact elliptical orbits
available in this limit. We then set $\dot{\beta}_{0}=\dot{\phi}_{0}=0$ so that all bulk
moduli fields have zero initial kinetic energy, while, of course, the initial value of
the Hubble parameter $\dot{\alpha}$ is given by the Friedmann equation~\eqref{Friedmann}.

\begin{figure}[htbp]
  \centering
   \includegraphics[width=6cm,angle=-90]{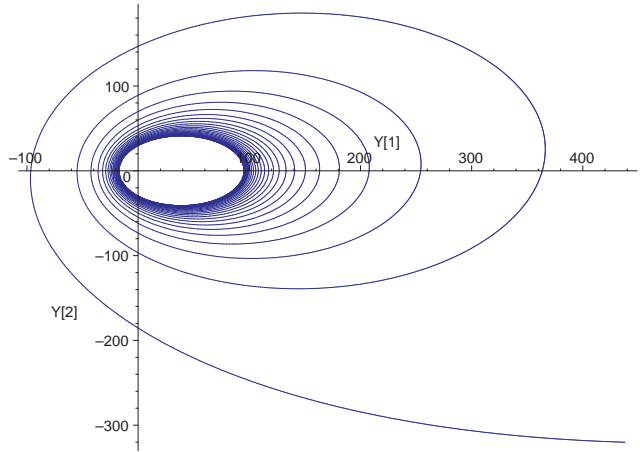}
   \caption{
   \textit{The $(Y^{1},Y^{2})$ phase-plane for $V=10^{9}$ and $\dot{\beta}_{0}=\dot{\phi}_{0}=0$, showing the
            diminishing effect of the radial force as the background evolves. If V is increased then successive
            antibrane orbits stay closer to each other, and to the origin, for a longer period of time. In the
            limit $V\rightarrow\infty$ they all degenerate into the exact ellipse trajectory visible at the centre.} }
        \label{fig1:orbit1}
\end{figure}

\begin{figure}[htbp]
  \centering
   \includegraphics[width=6cm,angle=-90]{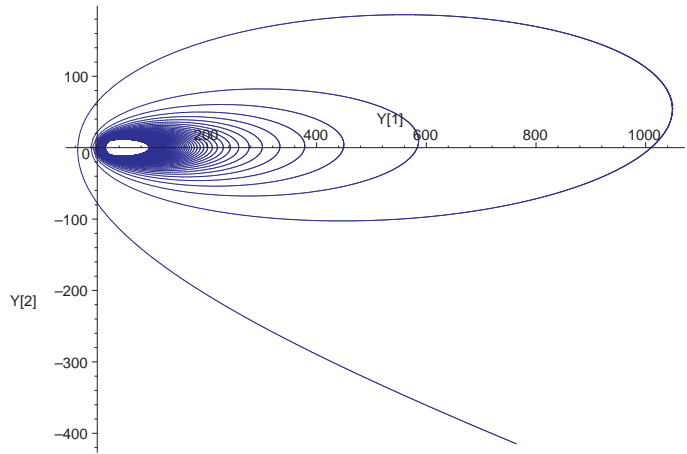}
   \caption{
   \textit{The $(Y^{1},Y^{2})$ phase-plane for the same initial conditions as in Fig.\ref{fig1:orbit1},
           but with k increased ten-fold. Thousands more revolutions are made before the antibrane approaches
          escape, at which point the compact geometry cannot be neglected and the analysis breaks down.} }
        \label{fig2:orbit2}
\end{figure}

The results are shown in Fig.~\ref{fig1:orbit1}. The necessarily small value of $1/V$ allows
the antibrane to orbit in elliptical fashion for many hundreds of revolutions, but these
orbits fail to close exactly due to a progressive diminution of the radial force. This leads
to the densely packed area of Fig.~\ref{fig1:orbit1}, which thins out as the evolution of
the background begins to dominate. Eventually the antibrane breaks free of the potential
well, and the motion becomes unbounded. We note that if $k_{0}$ is increased for fixed $V$,
such that $k_{0}=1000$ and $k=100$, then the escape point is pushed toward $R\sim1000$ and
the late-time analysis of the orbit can no longer be trusted (see Fig.~\ref{fig2:orbit2}).
Nonetheless, the early-time behaviour clearly shows the same trajectory mismatches, with the
antibrane gradually drifting away from the source.

This overall behaviour is easily explicable, once we notice that there is no longer a
conserved Runge-Lenz vector $\mathbf{A}$ after the moduli have been promoted to functions of
time. Recall that $\mathbf{A}$ usually lies in the plane of the orbit, is directed toward
the perihelion point, and has a magnitude equal to the eccentricity $e$ of the elliptical
trajectory. Its conservation thus guarantees the existence of non-precessing elliptical
orbits with fixed $e$. That it was conserved in the original branonium analysis can be
ascribed to the special, static form for $h$, which satisfied
\begin{align}\label{flux}
\frac{\partial{h}}{\partial{R}}\:R^{2}= \text{constant}
\end{align}
This simply states that, in the absence of external forces, the flux arriving at any
spherical surface surrounding a $1/R^{2}$ source is a constant, since the surface area over
which the field extends scales up as $R^{2}$. This property is obviously lost once the
time-dependent $h$ from \eqref{trueh} is inserted into \eqref{flux}, since then additional
moduli factors appear that cause the flux to vary. Indeed, from the antibrane's perspective
these variations in $\beta$ and $\phi$ modify the force it experiences \emph{as if} the
source charge is leaking away due to some external influence. The full system should,
therefore, be compared to a planet orbiting a mass-shedding star, for which it is evident
that no stable orbits are possible unless the mass of the star is stabilised. Thus, even
though the motion is still confined to a plane orthogonal to the conserved angular momentum
vector $\mathbf{L}$, there is a changing Runge-Lenz vector $\dot{\mathbf{A}} \neq 0$ that
rules out closed orbits in this plane.

\begin{figure}[htbp]
  \centering
   \includegraphics[width=6cm,angle=-90]{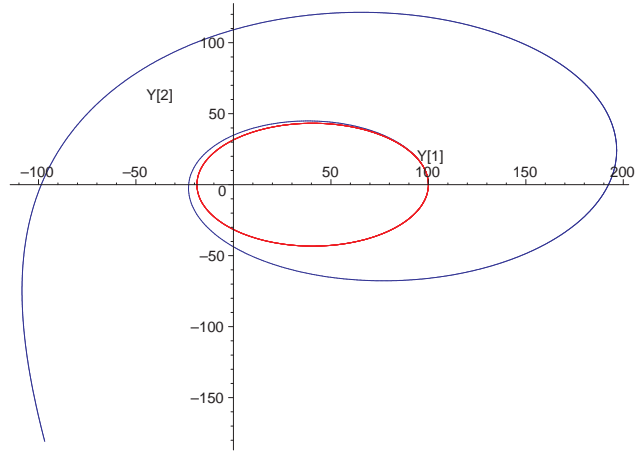}
   \caption{
   \textit{The $(Y^{1},Y^{2})$ phase-plane for $V\rightarrow \infty$, with a comparison between the exact
           ellipse $\dot{\beta}_{0}=\dot{\phi}_{0}=0$ and the initial conditions $\dot{\beta}_{0}=10^{-13}$,
           $\dot{\phi}_{0}=-10^{-12}$.   } }
       \label{fig3:orbit3}
\end{figure}

Given that there are no exactly closed orbits for finite $V$, we should account for this
cumulative mismatch in addition to the other instabilities such as radiation emission. We
reiterate that Fig.~\ref{fig1:orbit1} was produced for the gravitational moduli
being initially static. Without a specific stabilising potential present to arrange
this, such initial conditions must be considered quite special, and we should allow
$\dot{\beta}_{0},\dot{\phi}_{0}$ to take on non-zero values. In such cases
the antibrane will either spiral inward and annihilate, or quickly escape the binding
influence of the source, even in the limit of infinite transverse volume. As a simple
example of this, we can take $V \rightarrow \infty$ and
allocate $\beta$ and $\phi$ small initial kinetic energies, keeping all other parameters as
in Fig.~\ref{fig1:orbit1}. The results are shown in Fig.~\ref{fig3:orbit3}

\section*{Conclusion}

In this paper, we have derived the effective action for an anti D$p$-brane moving in the
background generated by a heavy D$p$ brane, taking into account the gravitational moduli
represented by the dilaton and the scale factor of the transverse space. For the case of a
finite transverse volume we find that these gravitational moduli cannot be consistently set
to constants once the anti D$p$-brane moves. This is due to non-trivial kinetic terms for
the anti-brane position moduli. In contrast, the gravitational moduli can be truncated off
consistently in the limit of infinite transverse volume, although the anti-brane motion will
still be affected as soon as the gravitational moduli evolve in time.

We have studied both the finite and infinite volume case in more detail focusing on
D6-branes. In a previous analysis~\cite{Burgess:2003qv} which neglected the gravitational
moduli, closed ``planetary'' type orbits were found for the anti D6-brane. Taking into
account the gravitational moduli we find in the finite volume case that orbits no longer
close and the anti D6-brane escapes from the central D6-brane after a number of revolutions.
This new feature is due to the time-evolution of the gravitational moduli, which is induced
by their cross-coupling with the brane-moduli despite being zero initially.

In the case of infinite transverse volume, the gravitational moduli can be set to constants
consistently leading to exactly closed orbits for the anti D6 brane. However, even small
initial velocities for the dilaton and the scale factor of the transverse space leads to
non-cyclic evolution of the anti D6-brane, which will either escape the central D6 brane or
spiral towards it, depending on initial conditions.

In this paper, we have focused on a simple class of models and have made use of
a number of approximations. For example, once the anti-brane escapes from the
central D-brane it will start probing the compact transverse space as well as
orientifold planes and possible other branes located elsewhere in the transverse
space. We have neglected such effects in deriving our effective theory and, therefore,
cannot analyse the long-term evolution of the anti-brane once it has escaped
from the central D-brane. It would be desirable to derive a more complete effective
theory and study this long-term evolution in detail.

\vspace{1cm}

\noindent
{\Large\bf Acknowledgements}\\
A.~L.~is supported by a PPARC Advanced Fellowship and J.~E.~ is supported
by a PPARC Studentship.

\bibliographystyle{plain}


\end{document}